\documentstyle[12pt,aaspp4]{article}
\begin{document}
\title{Low Surface Brightness Radio Structure in the Field \\
        of Gravitational Lens {\sf 0957+561}}

\author{I. M. Avruch, A. S. Cohen, J. Leh\'{a}r \altaffilmark{1}, 
        S. R. Conner, D. B. Haarsma, B. F. Burke}

\affil{Department of Physics, Research Laboratory of Electronics, \\
       Massachusetts Institute of Technology, Cambridge MA 02139}
\altaffiltext{1}{Center for Astrophysics, 60 Garden Street, Cambridge, MA
02138}

\begin{abstract}
We have produced deep radio maps of the double quasar {\sf 0957+561}
from multiple-epoch VLA observations.  To achieve high sensitivity to
extended structure we have re-reduced the best available 1.6~GHz
observations and have combined 5~GHz data from multiple array
configurations.  Regions of faint emission approximately
15\arcsec\,north and south of the radio source G are probably lobes
associated with the lensing galaxy.  An arc 5\arcsec\,to the east of G
may be a stretched image of emission in the background quasar's
environment.  1.4\arcsec\,southwest of G we detect a source that we
interpret as an image of emission from the quasar's western lobe,
which could provide a constraint on the slope of the gravitational
potential in the central region of the lens.  We explore the
consequences of these new constraints with simple lens models of the
system.

\end{abstract}

\section{Introduction}
Astrophysicists have anticipated the use of gravitational lensing as
an observational tool for 60 years (\cite{zwicky37a};
\cite{schneiderbook}), and in the case of the double quasar {\sf
0957+561} (\cite{walsh79}), after 18 years of study the promise is
closest to fulfillment.  If one knew the details of the
gravitational-lensing potential, and the time delay among the images
of flux-variable components, one could make an estimate, albeit
cosmology-dependent, of Hubble's constant H$_0$ (\cite{refsdal64.2}).

Efforts to measure the time delay in this system have converged
recently (417$\pm$3, \cite{kundic97}; 420$\pm$13, \cite{haarsma-th}).
However, models of the lensing potential have been less well
constrained (\cite{falco91}; \cite{kochanek91}; \cite{grogin96},
1996b) despite detailed observations of the cluster of galaxies
providing the lensing mass (\cite{young81}; \cite{angonin94};
\cite{fischer97}).  In an effort to produce a definitive radio map of
the object we undertook to re-reduce VLA\footnote{The VLA is part of
the National Radio Astronomy Observatory, which is operated by
Associated Universities, Inc. under co-operative agreement with the
National Science Foundation.} data gathered by the M.I.T. group,
discovering several new features in the field (\cite{avruch93}).  In
this letter we present improved maps and identify features that may be
useful as model constraints.
\section{Observations} 
The data sets from which results presented in this letter were
computed are listed in Table 1.  We first mapped a low resolution
$\lambda$6cm data set to identify any sources in the primary beam
whose side lobes would contaminate the field of interest; during
self-calibration of other data sets this emission was taken into
account.  For sensitivity to low surface brightness features we chose
the best extant $\lambda$18cm observation.  We have also combined five
$\lambda$6cm data sets from array configurations A, B, and C to
achieve more complete $(u,v)$ coverage and compensate for the reduced
brightness at $\lambda$6cm compared to $\lambda$18cm.  The fluxes of A
and B were roughly constant at these epochs. Using the {\tt AIPS}
software, the data sets were independently mapped and self-calibrated
following standard VLA reduction procedures (\cite{fomalont89}).  Each
data set was phase self-calibrated several times, followed by a single
amplitude self-calibration, provided that it reduced the map noise.
The individual data sets were then co-added in {\tt AIPS} and the
combined data were mapped and self-calibrated as above.  To produce
source-subtracted images, we use the model for compact emission that
the deconvolution algorithm creates in the form of {\tt CLEAN}
components, subtracting the model source from the visibility plane and
remapping. 

To the north and south of {\sf 0957+561} we have detected lobes of
emission, separated by about 30\arcsec.  The northern lobe (N) is more
compact, with a $\lambda$6cm flux of about 840 $\mu$Jy and spectral
index $\alpha^{\rm 18cm}_{\rm \phm{0}6cm} \sim$ $-$1.0
($S\propto\nu^\alpha$). The southern lobe (S) is extended, with a
total flux of about 1000 $\mu$Jy, $\alpha^{\rm 18cm}_{\rm \phm{0}6cm}
\sim$ $-$0.7.  These lobes may be associated with radio galaxy G, or
with the lensed quasar in the background.  To the east of the quasar
images A and B we have detected an arc of emission (R1).  The arc is
clearly resolved tangentially, with a peak flux of 1.27 mJy
beam$^{-1}$ at $\lambda$18cm and spectral index $\alpha^{\rm
18cm}_{\rm \phm{0}6cm} \sim$ $-$0.8.  In Figure~1 we present a radio
map of these features; B has been subtracted from the image in the
manner described above.

Galaxy G, the dominant contributor to the lensing potential, is
definitely extended to the east, southwest, and northwest.  To better
view the structure near G, we subtracted from the multi-epoch $(u,v)$
data all emission associated with the B quasar image, the BN component
(\cite{roberts85}), and G\@.  These structures were identified by
directly inspecting the {\tt CLEAN} components from the multi-epoch
map.  In Figure~2 the extension of G to the east we name GE, to the
northwest GN, to the northeast GNE, and the brightest component of the
arc-like structure to the southwest of G we call R2.  Table~2 presents
the positions and fluxes for these new components.

We are confident that these features are real.  N, S, and R1 have been
confirmed with detections by Harvanek~{\it et~al.\ }\,(1996); R1 and
perhaps GN have also been confirmed by Porcas~{\it et~al.\ }\,(1996).
The fainter features GE, GN, GNE, and R2 are visible in every
individual, reduced data set with sufficient resolution and
sensitivity, so it is unlikely that they are artifacts of calibration
or deconvolution.  On the other hand, detailed substructure such as
the double peaks of GNE is not significant, because with extended
sources {\tt CLEAN} produces spurious peaks on that scale
(\cite{briggs-th}).
\section{Discussion}
To illustrate our interpretation of these new VLA components, we used
the {\tt LENSMOD} software (\cite{lehar93}) to model the lensing mass
with a softened power-law potential (\cite{blandford87}).  The model
parameters were: the lens position ($\Delta\alpha$, $\Delta\delta$),
the critical radius ($b$), a core radius ($\theta_c$), the power index
$P$ ($P=1$ is isothermal, while $P=2$ is a Hubble profile), the
isodensity ellipticity ($e=1-\frac{\rm minor\;axis}{\rm
major\;axis}$), and the major axis orientation ($\phi$).  As
constraints we used the new HST quasar and G1 positions
(\cite{bernstein97}) and required that the quasar images have a
magnification ratio of 0.75$\pm$0.02 (\cite{schild90}).  We required
that any third image of the quasar near G be at least 30 times fainter
than B (as a 1$\sigma$ limit).  We also added constraints from the new
HST ``blobs'' and ``arc.''  We required that blob2 and blob3 be images
of each other, and that the two knots in the arc share a common
source.  Note that the HST arc is probably caused by the eastern end
of the same object that gives rise to blobs 2 and 3, and this could be
used to further constrain lens models. To account for the possibility
that the HST objects are at a different redshift than the quasar, we
added a uniform scale factor $Q_2$ to the deflection angles for those
components, as an extra model parameter.  The lens model parameters
were varied until the source plane position and magnitude differences
for each pair were minimized, with a resultant reduced $\chi^2$ for
the fit of 1.1.  The best fit model parameters are given in Table~3,
with uncertainties determined by varying the model parameters until
the reduced $\chi^2$ increased by 1.  Note that the $Q_2$ range
corresponds to HST component redshifts of $z_{\rm
HST}\approx1.3\pm0.1$ for an $\Omega=1$ cosmology, which is fully
consistent with the quasar and HST objects being at the same redshift.
Figure~3 shows the best fit model for $Q_2=1$, with components added
to show the modeled radio emission.  We do not attempt to account for
the VLBI structures (\cite{garrett94}) in this model, and thus make no
claims about the time delay or Hubble's constant based upon our model.

We interpret the component GE as the counter-image to the low surface
brightness tail of the quasar's western radio lobe E\@.  GE's peak
surface brightness and spectral index ($\alpha^{\rm 18cm}_{\rm
\phm{0}6cm} \sim$ $-$1.0) matches that of component E's northeastern
extension, so the brighter parts of the lobe are not multiply imaged.
The Bernstein {\it et~al.\ }\,(1996) HST blobs 2 and 3, almost
certainly multiple images of a background object, are very close to
the positions of GE and the northeast end of E; therefore we expect an
image of E near where we have found GE\@.  Because not all of E is
multiply imaged, the detailed structure of GE can yield strong
constraints on the central region of the lens: either the mass
distribution is non-singular, in which case GE comprises two merging
images of the eastern end of E, or, if the mass has a central
singularity, GE will have a sharp cusp at its western end.  High
resolution radio observations of GE may be able to distinguish these
two possibilities, or at least determine an upper limit on the size of
the central mass concentration in G.  This is also important because,
for a given lens mass, the potential near the quasar B~image is
generally deeper for singular models, yielding a longer predicted time
delay and thus a lower H$_0$ estimate.

The arc-like feature R1 may be a stretched image of background
emission.  As there is no clear counterpart to the west of G, it is
unlikely to be multiply imaged.  If the background source is circular,
the axial ratio of R1 yields a lower limit of about 5 for its
magnification.  Jones~{\it et~al.\ }(1993), in {\it Einstein} HRI
data, have detected an apparent x-ray arc about 3\arcsec\,northwest of
R1.  The positions are formally consistent, but seem unlikely to be
coincident judging from the relative positions of A and B\@.  An
association is not ruled out, however.  The authors claim the extended
x-ray source is an image of thermal emission from a cooling flow in
the cluster hosting the lensed quasar at z = 1.41.  There are examples
of diffuse non-thermal radio emission associated with x-ray-emitting
clusters (\cite{deiss97}), and in this case the lensing magnification
may have helped to make it observable.  Of course this emission could
be foreground; if G has radio lobes, N and S, it could as well have
jets.  R1 might be back flow from the lobe S, and GN might be a faint
jet feeding the lobe N\@.  GE is well explained as an image of the
quasar's E lobe, but it's not impossible for it to be the counter-jet
of GN, feeding lobe S\@.

The features R2, GN, and GNE are not readily explained by a lensing
hypothesis. R2 is in the position of the western half of the HST
arc, but all the models we have investigated would produce an
eastward extension of this arc which is not detected.  We could
appeal, {\it ad hoc}, to source size and spectral index morphology
causing the image to be unobservable.  The component GN should have a
brighter image 5\arcsec\,south of G, which is not seen, though we
could make the same appeal and note that it might be difficult to
separate visually from S\@.  GNE should also have a counter-image to the
south of G, which is not seen.  However, given the interpretation of
R1 as lensed, and the faintness of these features, it is not ruled out
that at least some of the emission is due to structure in the
background quasar's environment.

N and S are certainly not multiply imaged, but whether they are
foreground or background is less clear. They could be the radio lobes
of the galaxy G. At the lens redshift ($z = 0.36$, and assuming
$\Omega=1$, $h=0.75$) N and S would have a (projected) proper
separation of 120 kpc, and luminosities at 178~MHz of about 10$^{24}$
WHz$^{-1}$, typical values for low power, limb darkened radio
galaxies.  The optical classification of G as a cD galaxy, and the
fact that N and S are aligned within 30\arcdeg\, of its optical minor
axis are also consistent (\cite{miley80}).  N and S might be old lobes
of the background quasar, in which case the numbers are 170 kpc and
10$^{26}$WHz$^{-1}$, more appropriate for powerful, limb brightened
sources.  If N and S are associated with the quasar, the relatively
small lobe separation (56 kpc) and the high core-to-lobe flux ratio
($R=0.22$) suggest that the jet axis is moderately inclined towards
the line-of-sight (\cite{muxlow91}).  This inclination readily
explains the seemingly large rotation of the jet from the axis defined
by N and S to that defined by C and E.

The performance of the VLA at $\lambda$18cm has improved markedly
since 1980, and new observations should detect or exclude these
features with high significance.  We are aware of a very deep VLA
observation (\cite{harvanek96}) at $\lambda$18cm and $\lambda$3.6cm;
the longer wavelength data should be able to confirm GE, GN, and GNE,
and if GE is detected at $\lambda$3.6cm it may be possible to determine
whether the mass model is singular, or whether GE consists of two
merging images.

\acknowledgements This research has made use of NASA's Astrophysics
Data System Abstract Service, and the NASA/IPAC Extragalactic Database
(NED) which is operated by the Jet Propulsion Laboratory, California
Institute of Technology, under contract with the National Aeronautics
and Space Administration.  B.F.B is thankful for support from various
NSF grants, and J.L. is grateful for support from NSF grant
\#AST93-03527.

\clearpage

\begin{figure}
\plotone{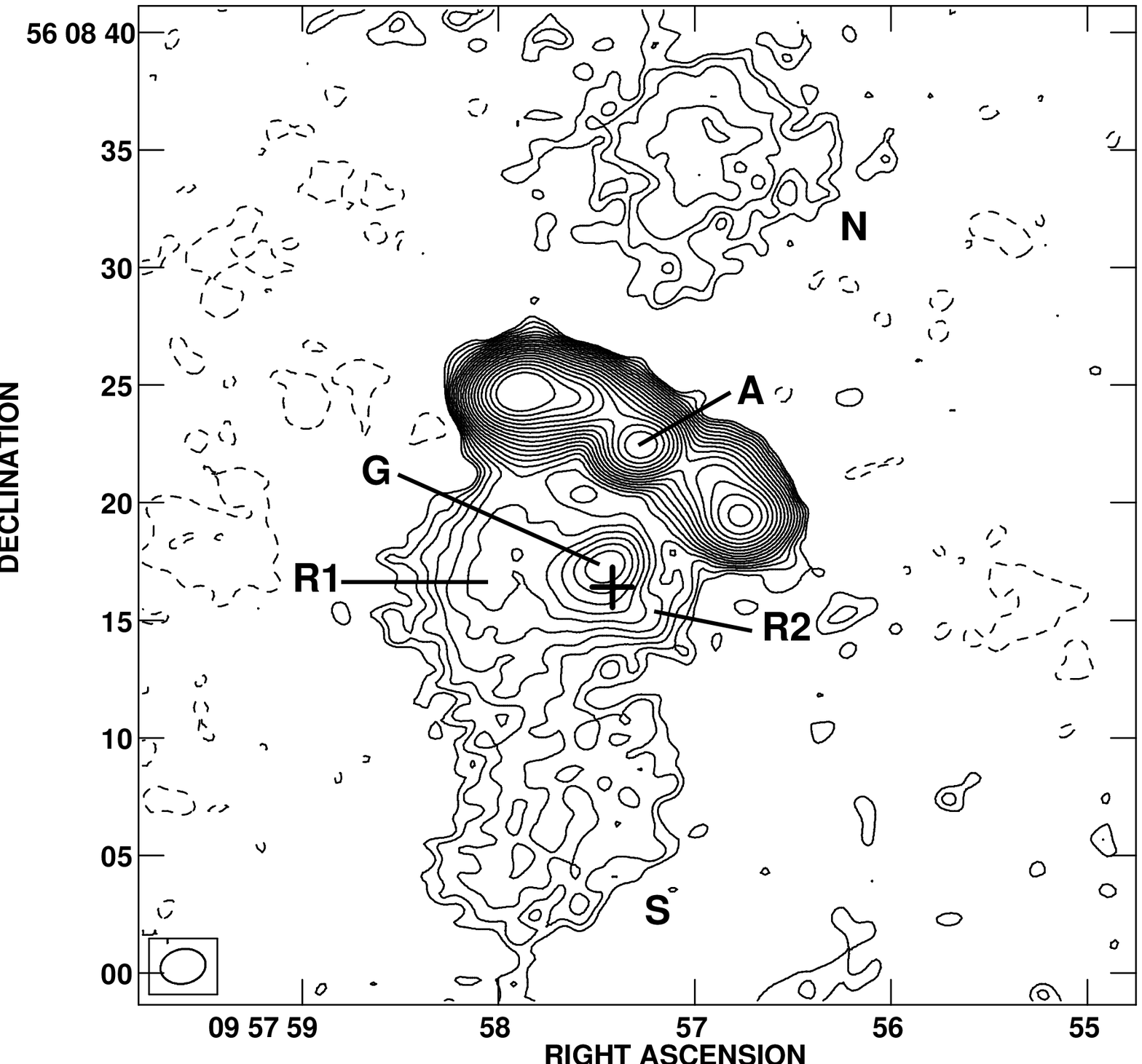}
\caption{Contour plot of $\lambda$18cm A array map of {\sf0957+561} on
1980 December 16.  The cross hair marks the position at which the quasar
B component has been subtracted from the map. The source just to the
north of B is G, the lensing galaxy.  Contour levels are $-$0.10\%,
0.10\%, 0.20\%, 0.28\%, 0.40\%, 0.57\%, 0.80\%, 1.13\%, 1.60\%,
2.26\%, 3.2\%, 4.53\%, 6.40\%, 9.05\%, 12.8\%, 18.1\%, 25.6\%, 36.2\%,
and 51.2\% of the peak intensity of 181 mJy beam$^{-1}$. The noise
level is 105$\mu$Jy beam$^{-1}$.  The box in the lower left shows the
beam FWHM ellipse.\label{fig1}}
\end{figure}

\begin{figure}
\plotone{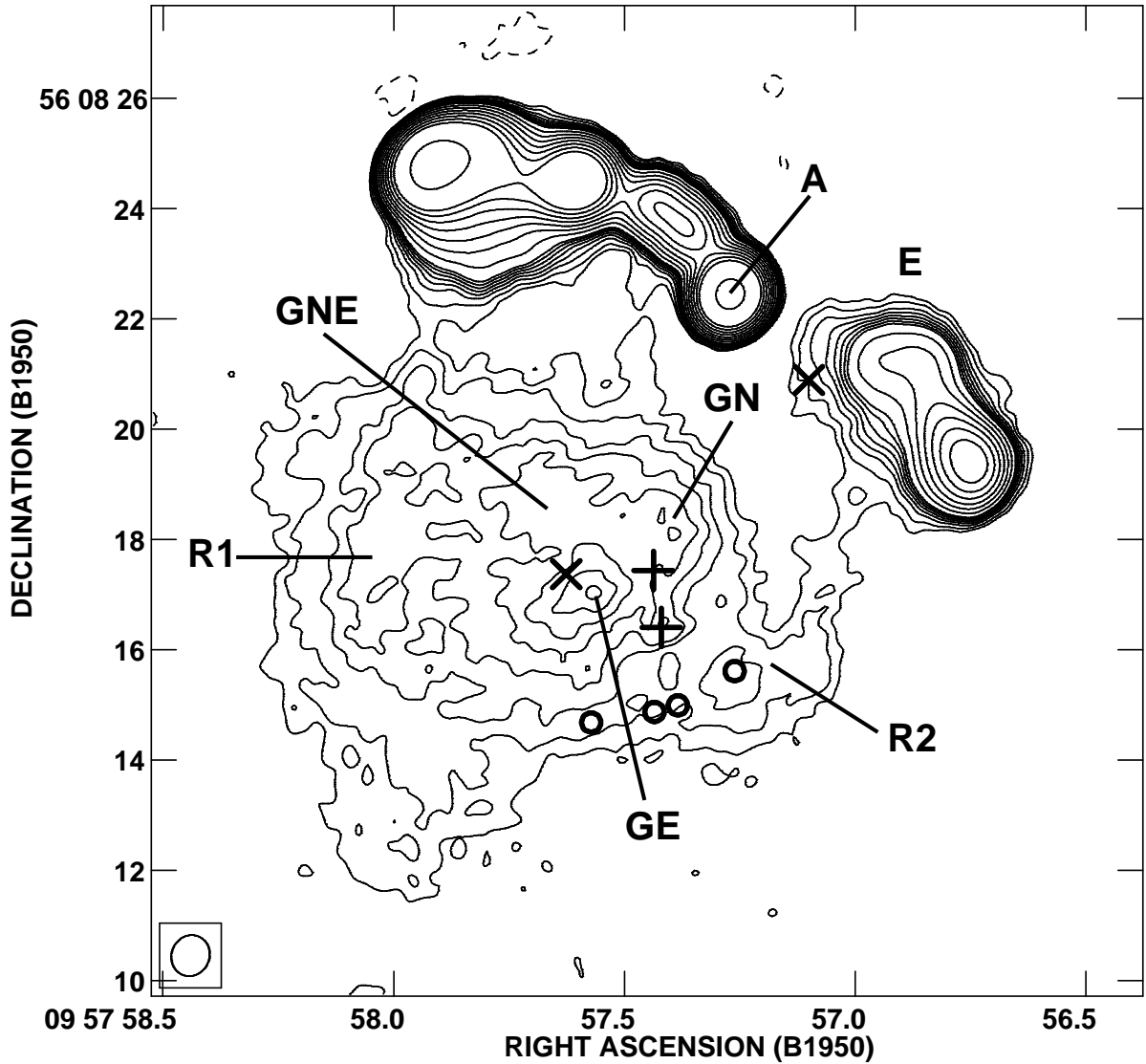}
\caption{Contour plot of $\lambda$6cm map of {\sf0957+561} from
co-added observations in A, B, and C arrays (data sets \#4 -- \#8,
Table 1). The cross hairs ($+$) are the positions from which models of
the components B (to the south) and G were subtracted.  The crosses
($\times$) are, east to west, the positions of HST components ``blob
2'' and ``blob 3.'' The circles are positions along the HST arc, the
outer two being the approximate extent and the inner two being ``knot
1'' and ``knot 2.''  Contour levels are $-$0.25\%, 0.25\%, 0.35\%,
0.50\%, 0.63\%, 0.75\%, 0.88\%, 1.00\%, 1.13\%, 1.60\%, 2.26\%,
3.20\%, 4.53\%, 6.40\%, 9.05\%, 12.8\%, and 51.2\% of the peak
intensity of 41.8 mJy beam$^{-1}$.  The noise level is 39$\mu$Jy
beam$^{-1}$.  The box in the lower left shows the beam FWHM
ellipse.\label{fig2}}
\end{figure}

\begin{figure}
\epsscale{0.5}
\plotone{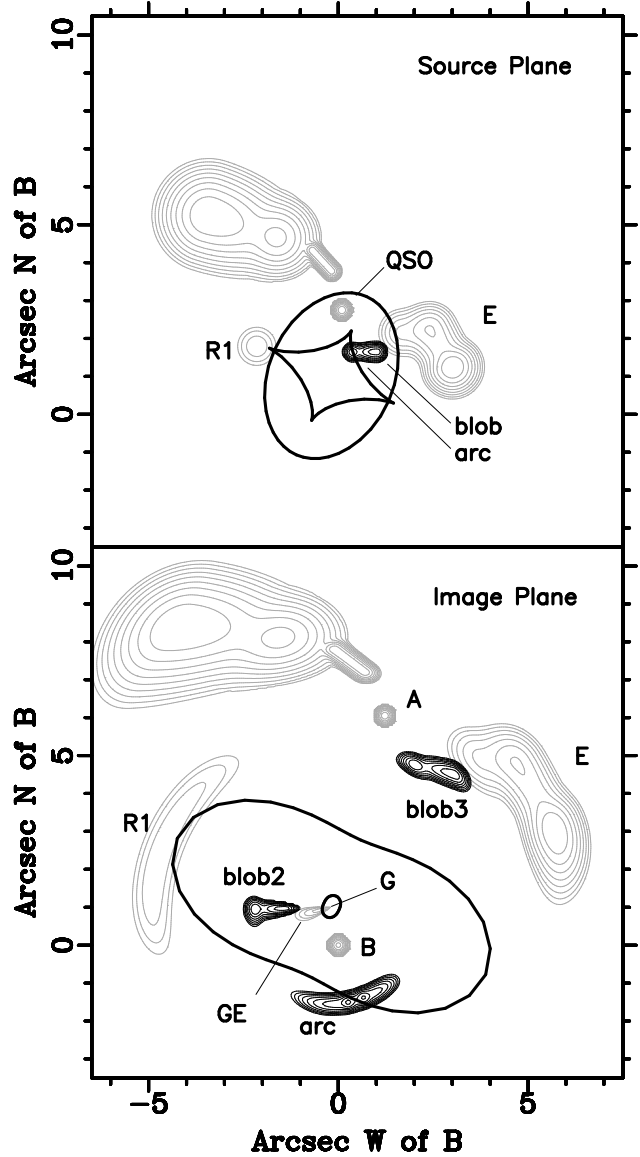}
\caption{Lens model constrained to the HST components, showing
disposition of radio components. The HST and radio components are
shown as dark and light contours, respectively.  The source plane
shows how the source would appear without lensing; the caustics
separate regions of multiple imaging.  The image plane shows the model
source seen through our lens model; the critical lines divide the
images.  The location of G is shown on the image plane, at the center
of the lens model.  Note that the HST arc is probably formed by the
eastern end of the source that yields the HST blobs.\label{fig3}}
\end{figure}

\clearpage

\begin{deluxetable}{cccccc}
\tablecolumns{6}
\footnotesize
\tablecaption{Archival VLA Data Presented in this Letter. \label{tbl-1}}
\tablewidth{0pt}
\tablehead{
\colhead{N$^o$} & \colhead{Obs. Date} & \colhead{$\lambda$ (cm)} & 
\colhead{Duration (hrs)\tablenotemark{a}} & \colhead{VLA Config.}}
\startdata
1 & 16DEC80 &    18 & 3.0 & A & \nl
2 & 21AUG90 & \phn6 & 0.8 & B & \nl    
3 & 17MAR88 & \phn6 & 0.9 & C & \nl
4 & 27SEP87 & \phn6 & 1.5 & A & \nl
5 & 20JUL87 & \phn6 & 0.8 & A & \nl
6 & 09DEC87 & \phn6 & 1.1 & B & \nl
7 & 02JUN85 & \phn6 & 1.2 & B & \nl
8 & 11JAN87 & \phn6 & 1.6 & C & \nl
\enddata
\tablenotetext{a}{The duration is the effective VLA observing time
devoted to {\sf 0957+561}, excluding time spent on calibration and
instrumental difficulties.}
\end{deluxetable}

\clearpage
\begin{deluxetable}{ccccc}
\tablecaption{Faint Emission Features Described in this Letter\label{tbl-2}}
\tablewidth{0pt}
\tablehead{
\colhead{{\small Feature}} & \colhead{{\small ($\Delta\alpha^{\arcsec}$,$\Delta\delta^{\arcsec}$)\tablenotemark{a}}} &
\colhead{{\small $\lambda$6cm Peak\tablenotemark{b}}} & \colhead{{\small $\lambda$6cm Flux}} & 
\colhead{{\small $\lambda$18cm Flux\tablenotemark{c}}} \\
\colhead{} & \colhead{} &
\colhead{{\small mJy beam$^{-1}$}} & \colhead{{\small mJy}} & \colhead{{\small mJy}}
}
\startdata
N  & ($-$4.5\phm{0},\phm{$-$}18.8\phm{0}) $\pm$  (0.5\phm{0},0.5\phm{0}) & 0.440$\pm$0.053 \tablenotemark{d} & 0.840$\pm$0.110 \tablenotemark{d} & 10.15$\pm$0.66 \nl
S  & (\phm{$-$}1.5\phm{0},$-$11.7\phm{0}) $\pm$  (0.5\phm{0},0.5\phm{0}) & 0.270$\pm$0.053 \tablenotemark{d} & 1.030$\pm$0.149 \tablenotemark{d} & \phn8.87$\pm$0.61 \nl
R1 & (\phm{$-$}5.38,\phm{$-$0}0.80)       $\pm$  (0.59,1.17)             & 0.366$\pm$0.043 \tablenotemark{e} & 3.646$\pm$0.181 \tablenotemark{e} & \phn6.61$\pm$0.31 \nl
R2 & ($-$1.28,\phm{0}$-$0.80)             $\pm$  (0.26,0.21)             & 0.250$\pm$0.039 \tablenotemark{f} & 2.270$\pm$0.159 \tablenotemark{f} & \nodata \nl
GE & (\phm{$-$}1.24,\phm{$-$}\phm{0}0.68) $\pm$  (0.04,0.04)             & 0.486$\pm$0.039 \tablenotemark{f} & 2.130$\pm$0.093 \tablenotemark{f} & \nodata \nl
GN & ($-$0.24,\phm{$-$0}1.68)             $\pm$  (0.12,0.24)             & 0.373$\pm$0.039 \tablenotemark{f} & 0.930$\pm$0.066 \tablenotemark{f} & \nodata \nl
GNE& (\phm{$-$}2.24,\phm{$-$0}2.08)       $\pm$  (0.34,0.22)             & 0.357$\pm$0.039 \tablenotemark{f} & 1.186$\pm$0.073 \tablenotemark{f} & \nodata \nl
\enddata
\tablenotetext{a}{Positions relative to B, $\alpha$$=$09$^{\rm
h}$57$^{\rm m}$57\fs42$\pm$0\fs01, $\delta$$=$56\arcdeg08\arcmin16\farcs40$\pm$0\farcs1 (B1950)}
\tablenotetext{b}{Flux uncertainties are based on the measured map
noise away from source emission.  For the components GE, GN, and GNE,
the error is dominated by the deconvolution algorithm, and the quoted
errors are likely underestimated.}
\tablenotetext{c}{$\lambda$18cm, beam FWHM ellipse 1.94\arcsec$\times$1.47\arcsec, map rms 105$\mu$Jy beam$^{-1}$}
\tablenotetext{d}{$\lambda$6cm, beam FWHM ellipse 3.51\arcsec$\times$2.57\arcsec, map rms 53$\mu$Jy beam$^{-1}$}
\tablenotetext{e}{$\lambda$6cm, beam FWHM ellipse 1.56\arcsec$\times$1.37\arcsec, map rms 43$\mu$Jy beam$^{-1}$}
\tablenotetext{f}{$\lambda$6cm, beam FWHM ellipse 0.75\arcsec$\times$0.69\arcsec, map rms 39$\mu$Jy beam$^{-1}$}
\end{deluxetable}
\clearpage

\begin{deluxetable}{cccc}
\tablecolumns{2}
\footnotesize
\tablecaption{Best Fit Lens Model Parameters \label{tbl-3}}
\tablehead{\colhead{Parameter} & \colhead{Best Fit Value}}
\startdata
$\Delta\alpha$\tablenotemark{a}   &      $+$0\farcs181$\pm$0\farcs001     \nl
$\Delta\delta$\tablenotemark{a}   &      $+$1\farcs019$\pm$0\farcs001     \nl
$b$                               &\phm{$+$1}2\farcs88$\pm$0\farcs021     \nl 
$\theta_c$                        &               ($<$0\farcs02)          \nl 
$P$                               &      0.88$^{\rm +0.04}_{\rm -0.02}$   \nl
$e$                               &     \phm{$+$}0.538$\pm$0.018          \nl
$\phi$\tablenotemark{b}           &         $+$65.2$\pm$0.5\phm{0}        \nl
$Q_2$                             &  \phm{$+$}1.03$\pm$0.02               \nl
\enddata
\tablenotetext{a}{offsets in right ascension and declination from the B quasar image.}
\tablenotetext{b}{orientation in degrees from north through east}

\end{deluxetable}


\begin{thebibliography}{}

\bibitem[Angonin-Willaime, Soucail, \&~Vanderriest 1994]{angonin94}
Angonin-Willaime, M.-C., Soucail, G., \&~Vanderriest, C. 1994, \aap,
291, 411

\bibitem[Avruch~{\it et~al.\ }1993]{avruch93}
Avruch, I. M., Conner, S. R., Becker, D. J., \&~Burke, B. F. 1993,
\baas, 25, 1403

\bibitem[Bernstein~{\it et~al.\ }1997]{bernstein97} Bernstein, G.,
Fischer, P., Tyson, J. A., \&~Rhee, G. 1997, \apjl, 483, L79

\bibitem[Blandford \&~Kochanek 1987]{blandford87} 
Blandford, R. D., \&~Kochanek, C. S. 1987, \apj, 321, 658

\bibitem[Briggs 1995]{briggs-th}
Briggs, D. S. 1995, Ph.D. thesis, New Mexico Institute of Mining and
Technology, Socorro

\bibitem[Deiss~{\it et~al.\ }1997]{deiss97}
Deiss, B. M., Reich, W., Lesch, H., \&~Wielebinski, R. 1997, \aap,
321, 55

\bibitem[Falco~{\it et~al.\ }1991]{falco91}
Falco, E. E., Gorenstein, M. V., \&~Shapiro, I. I.
1991, \apj, 372, 364

\bibitem[Fischer~{\it et~al.\ }1997]{fischer97}
Fischer, P., Bernstein, G., Rhee, G., \&~Tyson, J. A. 1997, \aj, 113, 521

\bibitem[Fomalont \&~Perley 1989]{fomalont89}
Fomalont, E. B., \&~Perley, R. A. 1989, in ASP Conference Series,
Vol.\ 6, Synthesis Imaging in Radio Astronomy, ed. R. A. Perley,
F. R. Schwab, \&~A. H. Bridle (San Francisco: Astronomical Society of
the Pacific), 83

\bibitem[Garrett~{\it et~al.\ }1994]{garrett94}
Garrett, M. A., Calder, R. J., Porcas, R. W., King, L. J., Walsh, D.,
\&~Wilkinson, P. N. 1994, \mnras, 270, 457

\bibitem[Grogin \&~Narayan 1996a]{grogin96}
Grogin, N. A. \&~Narayan, R. 1996, \apj, 464, 92

\bibitem[Grogin \&~Narayan 1996b]{grogin-err96}
Grogin, N. A. \&~Narayan, R. 1996, \apj, 473, 570

\bibitem[Haarsma 1997]{haarsma-th}
Haarsma, D., B. 1997, Ph.D. thesis, M.I.T., Cambridge

\bibitem[Harvanek~{\it et~al.\ }1996]{harvanek96}
Harvanek, M., Stocke, J., Tyson, T., \&~Rhee, G. 1996, \baas, 28, 843

\bibitem[Jones~{\it et~al.\ }1993]{jones93}
Jones, C., Stern, C., Falco, E., Forman, W., David, L., \&~Shapiro, I.
1993, \apj, 410, 21

\bibitem[Kochanek 1991]{kochanek91}
Kochanek, C. S. 1991, \apj, 382, 58

\bibitem[Kundi\'c~{\it et~al.\ }1997]{kundic97}
Kundi\'c, T., {\it et al.\ }1997, \apj, 482, 75

\bibitem[Leh\'ar~{\it et~al.\ }1993]{lehar93}
Leh\'ar, J., Langston, G. I., Silber, A., Lawrence, C. R., \&~Burke,
B. F. 1993, \aj, 105, 847

\bibitem[Miley 1980]{miley80}
Miley, G. 1980, in \araa, Vol.\ 18, ed. G. Burbidge, D. Layzer,
\&~J. G. Phillips (Palo Alto: Annual Review Inc.), 165

\bibitem[Muxlow \&~Garrington 1991]{muxlow91} 
Muxlow, T. W. B., \&~Garrington, S. T. 1991, in Beams and Jets in
Astrophysics, ed. P. A. Hughes (New York: Cambridge University Press),
52

\bibitem[Porcas~{\it et~al.\ }1996]{porcas96} 
Porcas, R. W., Patnaik, A. R., Muxlow, T. W. B., Garrett, M. A.,
Walsh, D. 1996, in Astrophysical Applications of Gravitational Lensing,
ed. C. S. Kochanek \& J. N. Hewitt (Boston: Kluwer Academic
Publishers), 349

\bibitem[Refsdal 1964]{refsdal64.2} 
Refsdal, S. 1964, \mnras, 128, 307

\bibitem[Roberts~{\it et~al.\ }1985]{roberts85}
Roberts, D. H., Greenfield, P. E., Hewitt, J. N., Burke, B. F.,
\&~Dupree, A. K. 1985, \apj, 293, 356

\bibitem[Schild \&~Smith 1990]{schild90}
Schild, R. E., \&~Smith, R. C. 1990, \aj, 101, 813

\bibitem[Schneider, Ehlers, \&~Falco 1992]{schneiderbook}
Schneider, P., Ehlers, J., \&~Falco, E. E., 1992 Gravitational Lenses
(New York: Springer-Verlag)

\bibitem[Walsh, Carswell, \&~Weymann 1979]{walsh79}
Walsh, D., Carswell, R. F., \&~Weymann, R. J.  1979, \nat, 279, 381

\bibitem[Young~{\it et~al.\ }1981]{young81}
Young, P., Gunn, J. E., Kristian, J., Oke, J.B., \&~Westphal,
J. A. 1981, \apj, 244, 736

\bibitem[Zwicky 1937]{zwicky37a}
Zwicky, F. 1937, Phys. Rev., 51, 290
\end{thebibliography}
\end{document}